\documentclass[a4paper,10pt]{article}
\usepackage[utf8]{inputenc}
\usepackage{amsmath}
\usepackage{float}
\usepackage{hyperref}
\usepackage[bf]{caption}
\usepackage{graphicx}
\usepackage{subfig}
\usepackage{color}
\usepackage{amssymb}
\usepackage{authblk}
\usepackage[all]{xy}

\numberwithin{equation}{section}

\title{
\vskip 5 mm
Implication of U-duality for black branes in M/string theory}
\author{Samrat Bhowmick\thanks{email: samrat@physics.iitm.ac.in}}
\affil{\normalsize\it Department of Physics, \authorcr 
\it Indian Institute of Technology Madras, \authorcr \it Chennai 600036, India.}

\begin{document}
\maketitle

\begin{abstract}
\noindent
U-duality symmetry of M-theory and S and T-duality of string 
theory can be used to study various black branes solutions. 
We explore some aspect of this idea here. This symmetry can be used to get relations
among various components of the metric of the black brane. These relations in turn 
give relations among various components of the energy-momentum tensor. We show that, 
using these relations, without knowing the explicit form of form fields, we can get 
the black brane solutions. These features were studied previously in the context of M-theory.
Here we extensively studied them in string theory (type II supergravity). We also show 
that this formulation works for exotic branes. We give an example of a time-dependent 
system where this method is essential.
\end{abstract}

\newpage

\section{Introduction}
Black Holes are widely studied object in string theory. Various properties of a 
class
of black holes have been successfully described using mutually
BPS intersecting configurations of string/M-theory branes.
Mutually BPS intersecting configurations mean that two or more branes intersect
in such a way that they preserve half of the supersymmetries as the single brane does;
for
example, in M-theory, two stacks of 2-branes intersect at a
point; two stacks of 5-branes intersect along three common
spatial directions; a stack of 2-branes intersects a stack of 
5-branes along one common spatial direction; waves, if present,
are parallel to a common intersection direction; and each stack of
branes is smeared uniformly along the other brane directions.
See \cite{Tseytlin:1996bh,Tseytlin:1996hi} for more details and for
other such string/M-theory configurations.
Black hole entropies are calculated from
counting excitations of such configurations, and Hawking
radiation is calculated from interactions between them. 

Such brane configurations consist of only
branes but no antibranes in the extremal limit. In the near extremal limit, they
consist of a small number of antibranes also. It is the
interaction between branes and antibranes that gives rise to
Hawking radiation. String theory calculations are tractable and
match those of Bekenstein and Hawking in the extremal and near
extremal limits. 
In the full non-extremal limit, it is difficult to control string theory 
calculation.
However, even in the far-extremal limit, black hole dynamics is
expected to be described by mutually BPS intersecting brane
configurations where they now consist of branes, antibranes, and
other excitations living on them, all at non-zero temperature
and in dynamical equilibrium with each other \cite{Horowitz:1996ay} --
\cite{Bhowmick:2007tt}. From now on, for the sake of brevity, we will refer to such 
far-extremal configurations also as brane configurations, even though
they may now consist of branes and antibranes, left moving and
right moving waves and other excitations.

The entropy $S$ of $N$ stacks of mutually BPS intersecting brane
configurations, in the limit where $S \gg 1 \;$, is expected to
be given by
\begin{equation}\label{sn}
S \sim \prod_I \sqrt{n_I + \bar{n}_I} 
\sim {\cal E}^{\frac{N}{2}} \;,
\end{equation}
where $n_I$ and $\bar{n}_I \;$, $I = 1, \cdots, N \;$, denote
the numbers of branes and antibranes of $I^{th}$ type, ${\cal
E}$ is the total energy, and the second expression applies for
the charge neutral case where $n_I = \bar{n}_I$ for all $I$.
When $N \ge 3$, this system describes a proper black hole with
horizon. 
The proof for this expression is given by comparing it in
various limits with the entropy of the corresponding black holes
\cite{Horowitz:1996ay, Horowitz:1996ac}, see also \cite{Danielsson:2001xe} 
-- \cite{Mathur:2008ez}. For $N \le 4
\;$ and when other calculable factors omitted here are restored,
this expression matches that for the corresponding black holes
in the extremal and near extremal limit and, in the models based
on that of Danielsson et al. \cite{Danielsson:2001xe}, matches up to a numerical
factor in the far-extremal limit \cite{Horowitz:1996ay} -- \cite{Mathur:2008ez}
also.

Note that, in the limit of large ${\cal E} $, the entropy
$S({\cal E})$ is $\ll {\cal E}$ for radiation in a finite volume
and is $ \sim {\cal E}$ for strings in the Hagedorn regime. In
comparison, the entropy given in (\ref{sn}) is much larger when
$N > 2 \;$. This is because, the branes, in the mutually BPS
intersecting configurations, form bound states, become
fractional, and support very low energy excitations that lead
to a large entropy. Thus, for a given energy, such brane
configurations are highly entropic.

Another consequence of fractional branes is ``fuzz ball'' proposal. 
According to the fuzz ball picture for black holes
\cite{Mathur:2005zp,Mathur:2005ai}, the fractional branes arising from the bound states
formed by intersecting brane configurations have non-trivial
transverse spatial extensions due to quantum dynamics. The size
of their transverse extent is of the order of Schwarzschild
radius of the black holes. Therefore, essentially, the region
inside the `horizon' of the black hole is not empty but is
filled with fuzz ball whose fuzz arise from the quantum dynamics
of fractional strings/branes.

This fuzz ball picture of black hole was extended to study
early universe \cite{Chowdhury:2006pk,Mathur:2008ez}. This early universe was further studied
in \cite{Kalyana Rama:2007ft} -- \cite{Bhowmick:2010dd} (See also 
\cite{Bhowmick:2012dc}.) using 
U-duality symmetry of string/M-theory. 

U-duality is a symmetry of M-theory which consists of T-duality, S-duality of 
string theory and 
dimensional reduction and dimension upliftment. In certain cases of supergravity
solutions, these symmetries can be used to get relations among various metric 
components. These relations can be used to get relations among various 
components of the energy-momentum tensor. This method was proposed in 
\cite{Kalyana Rama:2007ft} and further discussed in 
\cite{Bhowmick:2008cq,Bhowmick:2010dd,Bhowmick:2012dc}. 
We review this idea following mainly \cite{Bhowmick:2010dd,Bhowmick:2012dc}. 
We show here this duality technique works for black holes as well.
Our main goal in this paper is to show this method works for string theory
black holes as well.

In this work, we also 
studied whether this duality method works
for exotic black brane solutions. 
Exotic branes are co-dimension 2 objects. 
It is known from study of low dimensional string/M-theory that there are certain 
``exotic'' particle whose higher dimensional origin can not be usual brane. 
These are called exotic states and their higher dimensional branes are called
exotic branes \cite{Elitzur:1997zn} --
\cite{deBoer:2012ma}.
These branes solutions have certain non-geometric features
\cite{deBoer:2010ud,deBoer:2012ma}.
One should note here that for such branes, metric components are not only function of
radial coordinate but also function of the angular coordinate. 
This type of non-geometric backgrounds is called
T-fold or more generally U-fold \cite{Hull:2004in}.
We discuss here such co-dimension 2 branes  
and try to see what the U-duality relations come out among various metric components 
and hence among the components of the energy-momentum tensor.

The relations among various 
components of the energy-momentum tensor, found from duality relations, are the keys for the cosmological
model discussed in \cite{Kalyana Rama:2007ft} -- \cite{Bhowmick:2012dc}. 
A similar construction was used to get star like solutions
of M-theory in \cite{Rama:2011xz,Rama:2013jfa,Rama:2013vya}.
In case of black branes, the action is known explicitly, but in case of said cosmological
model fields are not known. There, duality relations play a most important role. A similar
model of an early universe in string theory may become useful because in string theory
interaction of various branes and strings are more tractable than that of M-theory.

This paper is organized as follows. Here we give a brief review of a part of work
of \cite{Kalyana Rama:2007ft} -- \cite{Bhowmick:2012dc} in section \ref{sec:Mbh}. 
Where we show U duality relations for black holes in M-theory. In section \ref{sec:Exbh}
we show similar relations exist for exotic M brane solutions. In section \ref{sec:Stringbh}
similar relations for string theory branes are studied using S and T-duality. 
We discuss possible applications in section \ref{sec:TD} to conclude the paper.

\section{General Black Brane solutions} \label{sec:Mbh}

In black brane solutions, $T_{A B}$ is obtained from the action of
higher form gauge fields. 
With a suitable ansatz for the metric,
equations of motion can be solved to obtain a black hole
solutions. 

To explain the process consider 11-dimensional supergravity
action. The bosonic part of the action is
 \begin{equation}
 \label{Mact}
 S = \frac{1}{16 \pi G_{11}} \int d^{11}x \, \sqrt{-g} \left( R
   - \frac{1}{2\times 4!} F_4^2 \right)\; ,
\end{equation}
where $F_4$ is a 4-form field strength, $F_4 = d C_3$.

The 
theory given by (\ref{Mact}) contains a 2 dimensional and a 5 dimensional 
objects $M2$ and $M5$ branes.
$M2$ branes are electrically charged and $M5$ branes are magnetically charged 
under $F_4$.
Energy-momentum tensor, $T_{AB}$, for this matter field, $F_4$, is given by
\begin{equation}
 \label{bhTAB}
 T_{AB} = \frac{1}{48}  \left[4 F_{AMNP} \, F_B{}^{MNP}
          -\frac{1}{2} g_{AB} \, F_4^2 \right] \; .
\end{equation}

This theory contains solutions with
solitonic objects.  
These black holes are actually made of stack of $M2$ or $M5$ branes
or mutually BPS intersecting combinations of them. With a suitable ansatz for metric 
and fields, Einstein equations can be solved to obtain black hole solutions.

To get the solutions, let the spacetime coordinates be $x^A = (r,
x^\alpha) \;$ where $x^\alpha = (x^0, x^i, \theta^a) $ with
$x^0 = t \;$, $i = 1, \cdots, q \;$, $a = 1, \cdots, m $, and
$q + m = 9 \;$. The $x^i$ directions may be taken to be
toroidal, some or all of which are wrapped by branes, and
$\theta^a$s are coordinates for an $m$ dimensional space of
constant curvature given by $\epsilon = \pm 1$ or $0 \;$.
The
metric and brane fields depend only on the $r$ coordinate, 
defined by $r^2=\sum_{\alpha=q+1}^{q+m}(x^\alpha)^2$. We write
the line element $d s$, in an obvious notation, as
\begin{equation}\label{dsbh}
d s^2 = - e^{2 \lambda^0} d t^2 
+ \sum_i^q e^{2 \lambda^i} (d x^i)^2 + e^{2 \lambda} d r^2 
+ e^{2 \sigma} d \Omega^2_{m, \epsilon} \; \; .
\end{equation} 
Black hole solutions are given by $\epsilon = +1$. 
But the
analysis is true for any maximally symmetric non-compact space.

The independent non-vanishing components of $T^A_{\; \; B} \;$
are given by $T^r{}_r = P_R $ and $T^\alpha{}_{\alpha}
= P_\alpha$, where $\alpha = (0, i, a) \;$. These components
can be calculated explicitly using the action $S_{br} \;$. For
example, for an electric $p$-brane along $(x^1, \cdots, x^p) \;$
directions, they are given by (see equation (\ref{bhTAB}))
\begin{equation}\label{electric}
P_0 = P_\parallel = - P_\perp = - P_a = P_R
= \frac{1}{4} \; F_{0 1 \cdots p r} \; 
F^{0 1 \cdots p r} \;,
\end{equation} 
where $P_\parallel \; = P_i$ for $i = 1, \cdots, p \;$, $P_\perp
\; = P_i$ for $i = p + 1, \cdots, q \;$, and note that $ P_R\;$ is
negative. For mutually BPS $N$ intersecting brane
configurations, it turns out 
\cite{Aref'eva:1996fw} -- \cite{Gal'tsov:2005vf} that
the respective energy-momentum tensors $T^A{}_{B}$ and
$T^A{}_{B (I)}$ obey conservation equations separately. 
\begin{equation}\label{tabI}
T^A{}_B = \sum_I T^A{}_{ B (I)} 
\; \; , \; \; \; 
\sum_A \nabla_A  T^A{}_{B (I)} = 0 \; \; . 
\end{equation}
In case of configurations with non-BPS intersection, this doesn't
hold. We show this by an example in the appendix.

Equations of motion may now
be written as
\begin{eqnarray}
\Lambda_r^2 - \sum_\alpha (\lambda^\alpha_r)^2 & = & 
2  P_R + \epsilon \; m (m - 1) e^{- 2 \sigma} \;, \label{t21bh} \\
e^{-2\lambda}\left[\lambda^\alpha_{r r} + (\Lambda_r-\lambda_r) \lambda^\alpha_r\right] & = & 
- \; P_\alpha + \frac{1}{9} \left(P_R + \sum_\beta P_\beta \right) 
\nonumber \\
& & + \; \epsilon \; (m - 1) e^{- 2 \sigma} \; \delta^{\alpha a}
\label{t22bh} \;,\\
{P_{R}}_r + P_R \Lambda_r - \sum_\alpha P_\alpha \lambda^\alpha_r & = & 0 \;,
\label{t23bh}
\end{eqnarray}
where $\Lambda = \sum_\alpha \lambda^\alpha = \lambda^0 + \sum_i
\lambda^i + m \sigma \;$ and the subscripts $r$ denote
$r$-derivatives. 
\subsection{Example} \label{sec:bbsl}
In this section, we give example of black $M2$, $M5$ brane solutions and their intersecting
configurations. We see that relations among scale factors exist.

\subsubsection*{$M2$ Branes} \label{sec:2}
Consider a stack of $M2$ branes along $(x^1,x^2)$ directions. $x^1$ and $x^2$ 
are taken to be compact. $x^3$ and $x^4$ are also taken to be compact.
In this case, the solution is known, 
line element of this solution can be taken to be 
\begin{equation}
\label{bbraneanz}
 ds^2 = -e^{2\lambda^0(r)}\,dt^2+\sum_{i=1}^4 e^{2\lambda^i(r)}\,(dx^i)^2
      + e^{2\lambda(r)} \left(dr^2+r^2 d\Omega^2_5 \right) \;.
\end{equation}
Ansatz for field, $C_{MNP}$ is $C_{012} = f(r)$ which gives $F_{012r} = \frac{df(r)}{dr}$.
This actually means $M2$ branes are electrically charged under the field $F_4$.
Energy momentum tensor is given by
\begin{equation}
 \label{emt2}
 T^0{}_0 = T^\parallel{}_\parallel = -T^\perp{}_\perp = T^r{}_r = 
 -T^a{}_a \;,
\end{equation}
where indices $\parallel$ and $\perp$ indicate parallel and
perpendicular to the brane directions
and $a$ indicates directions in $\Omega_5$ respectively. 
One can see from equations (\ref{emt2}) and (\ref{t22bh}) that the 
constraining relation
among scale factors, mentioned before turns out to be
\begin{equation}
 \label{sf2}
 \lambda^0 = \lambda^\parallel = - 2\lambda^\perp \;.
\end{equation}

\subsubsection*{BPS Intersection of 2 Sets of $M2$ Branes} \label{sec:22'}

Now consider 2 sets of intersecting $M2$ branes along $(x^1,x^2)$ and 
$(x^3,x^4)$. We denote first set by 2 and the second set by $2'$. Black brane 
solution of intersecting branes was first identified in 
\cite{Papadopoulos:1996uq}, then
many solutions were quickly constructed and governing rules of their existence 
were studied. 
This intersecting configuration follows BPS rules. For these
configurations our line element is 
\begin{equation*}
 ds^2 = -e^{2\lambda^0(r)}\,dt^2+\sum_{i=1}^4 e^{2\lambda^i(r)}\,(dx^i)^2
      + e^{2\lambda(r)} \left(dr^2+r^2 d\Omega^2_5 \right) \;.
\end{equation*}
Nevertheless, we have now two sets of electrically charged branes, so  
we have two non-zero components of the gauge field, $C_{012}(r)$ and $C_{034}(r)$ and their cyclic
permutations.

One can see easily from explicit expression of energy-momentum tensor
that, total energy-momentum tensor is
just the sum of energy-momentum tensors of individual brane configurations, $T^A{}_B = \sum_I \; T^A{}_{B\,(I)}$.
One can see that conservation equation is 
satisfied for total energy-momentum tensor as well as individual energy
momentum tensors. 
Just like in previous subsection, relations among scale factors 
come out, namely,
\begin{eqnarray}
 \label{sf22}
 \lambda^1 &=& \lambda^2 \nonumber \\
 \lambda^3 &=& \lambda^4 \nonumber \\ 
 2\lambda^1+2\lambda^3+\lambda^0 &=& 0\;.
\end{eqnarray}

\subsubsection*{$M5$ Branes} \label{sec:5}
In case of $M5$ brane, just like $M2$ brane case, metric ansatz is taken in the 
same form, except now 5 of the 10 spacelike dimensions 
$(x^1,x^2,x^3,x^4,x^5)$ are compact and $M5$ branes wrap them. In general
we may take some of the other directions are also compact. In that case they 
will be treated as directions perpendicular to branes and will be in 
same footing as directions of $\Omega_4$.
Metric ansatz is taken to be
\begin{equation}
 ds^2 = -e^{2\lambda^0(r)} dt^2+\sum_{i=1}^5 e^{\lambda^i} (dx^i)^2
      + e^{2\lambda(r)} \left(dr^2+r^2 d\Omega^2_4 \right) \;,
\end{equation}
where now $r^2 = \sum_{i=6}^{10} (x^i)^2$.
$M5$ branes are magnetically charged under the gauge field $F_4$. $C_{NPQ}$ is 
$C_{NPQ} = \frac{1}{4} \epsilon_{012345rMNPQ}
 f(r) x^M$. So $F_4$ takes non-zero value only when ${M,N,P,Q} \in \Omega_4$.
With this ansatz energy-momentum tensor turns out to be
\begin{equation}
 \label{emt5}
 T^0{}_0 = T^\parallel{}_\parallel =  T^r{}_r = 
 -T^a{}_a = -\frac{1}{4}\; ,
\end{equation}
where index $\parallel$ indicates parallel to brane directions
and $a$ indicates directions in $\Omega_4$ respectively.
Equations (\ref{emt5}) and equations of motion imply relations among
scale factors. These equations are same as (\ref{sf2}).
\begin{equation}
 \label{sf5}
 2 \lambda^0 = 2 \lambda^\parallel = - \lambda^\perp \;.
\end{equation}

\subsubsection*{BPS Intersection of $M2$ Branes and $M5$ Branes} \label{sec:25}
In this subsection, we give an example of BPS intersecting configuration of
a stack of  $M2$ branes is stretched along 
$(x^1,x^2)$ and that of $M5$ branes is stretched along 
$(x^1,x^3,x^4,x^5,x^6)$. All these $(x^1 \cdots x^6)$ are compact as before, and
the system is localised in common transverse space $(x^7 \cdots x^{10})$.
Again ansatz for black brane metric is similar to previous cases. It
is taken in the form
\begin{equation}
 \label{bh25anz}
 ds^2 = -e^{2\lambda^0(r)} dt^2+\sum_{i=1}^6 e^{\lambda^i} (dx^i)^2
      + e^{2\lambda(r)} \left(dr^2+r^2 d\Omega^2_3 \right) \;.
\end{equation}
Here $r^2 = \sum_{i=7}^{10}(x^i)^2$.
Under 4-form gauge field $M2$ branes are electrically charged and $M5$
branes are charged magnetically. Here non-zero components of gauge potential
are $C_{012}(r)$ and $C_{NPQ}(r,x^M)$, where ${M,N,P,Q} \in \Omega_3$.
The non-zero components of
energy-momentum tensor for this
set of fields are $T^i{}_i$, $i=0,1,\cdots,6$, $ T^r{}_r$ and $T^a{}_a$, $a \in \Omega_3$.
Use of explicit expression of above components 
and the equations of motion imply constraining relations, like
before, among scale factors.
\begin{eqnarray}
 \label{sf25}
 \lambda^0 &=& \lambda^1 \nonumber \\
 \lambda^3 &=& \lambda^4 \;=\; \lambda^5 \;=\; \lambda^6 \nonumber \\
 \lambda^2+2\lambda^3 &=& 0\;.
\end{eqnarray}

\subsection{U Duality Relations In M-Theory} \label{sec:Ud}
We now describe the relations which follow from U duality
symmetries, involving chains of dimensional reduction and
uplifting and T and S dualities of string theory.
These relations were found and used in case of cosmological solution
previously.
To explain the concept let us 
consider a solution of the form
\begin{equation}\label{dsgen11}
 ds_{11}^2 = - e^{2\lambda^0}dt^2 +
 \sum_{\mu=1}^{q} e^{2 \lambda^\mu} (d x^\mu)^2 + e^{2 \lambda} d r^2 
+ e^{2 \sigma} d \Omega^2_{m, \epsilon}\;,
\end{equation}
where we assume for $\mu = i, j, k$; $x^i$s are compact and Killing directions. 
That is $\lambda^\mu = \lambda^\mu(t,X)$, where $X$ 
includes space like coordinates except $x^i, x^j$ and $x^k$.
Let $\downarrow_k$ and
$\uparrow_k$ denote dimensional reduction and uplifting along
$k^{th}$ direction between M-theory and type IIA string theory.
To apply $\downarrow_{k}$ on the metric given in (\ref{dsgen11})
we write $ds_{11}$ as
\begin{equation}
  \label{downk}
  ds_{11}^2 = e^{-\frac{2}{3} \phi} \; ds_{10}^2 
  + e^{\frac{4}{3} \phi} \, (dx^{k})^2 \;,
\end{equation}
where $ds_{10}$ is 10 dimensional line element of type IIA theory. 
Type IIA string theory metric is given by
\begin{equation}
 \label{dsIIA}
 ds_{10}^2 = - e^{2\lambda'^{0}}dt^2 
 + \sum_{\mu \ne k} e^{2 \lambda'^\mu} (d x^{\mu})^2 
 + e^{2 \lambda'} d r^2 
+ e^{2 \sigma'} d \Omega^2_{m, \epsilon}\;.
\end{equation}
$\phi$ is dilaton and is independent of $x^i, x^j$ and $x^k$. It is
function of $(t,X)$ only.
If we integrate over $x^{k}$ with above metric we will get type 
IIA supergravity action.
Here, comparing equations (\ref{dsgen11}), (\ref{downk}) and 
(\ref{dsIIA}) one can see
$\phi$ and $\lambda'^\mu$ are given by
\begin{eqnarray}
 \label{down_k}
 \phi &=& \frac{3}{2} \lambda^{k}  \nonumber \\
 \lambda'^\mu &=& \lambda^\mu + \frac{1}{3} \phi
 = \lambda^\mu +\frac{1}{2} \lambda^k \nonumber \\
 \lambda' &=& \lambda +\frac{1}{2} \lambda^k \nonumber \\
 \sigma' &=& \sigma +\frac{1}{2} \lambda^k \nonumber \\
\end{eqnarray}

In string theory, application of T-duality along a compact direction
converts type IIA theory to type IIB theory and back. 
It also converts a $Dp$ brane to $D(p-1)$ or $D(p+1)$ branes depending on
whether T-duality is applied along the brane or perpendicular to the brane 
respectively.
Applying this transformation generates a new solution. 
We denote T-duality operation along $i^{th}$ direction by $T_i$.

Applying a T-duality along say, $x^j$, which we denote by $T_j$, generates
a new solution, given by
\begin{eqnarray}
 {ds'}_{10}^{2} &=& - e^{2\lambda'^{0}}dt^2 
 + \sum_{\mu \ne \{j,k\} } 
 e^{2 \lambda'^\mu} (dx^\mu)^2 + e^{-2 \lambda'^j} (dx^j)^2
 + e^{2 \lambda'} d r^2 
+ e^{2 \sigma'} d \Omega^2_{m, \epsilon} \;,
\nonumber \\
 \phi' &=& \phi - \lambda'^j = \lambda^k - \lambda^j \;,
\end{eqnarray}
where equation (\ref{down_k}) has been used. Note that metric along $x^j$, 
$g_{jj}$ goes to $(g_{jj})^{-1}$. 
This solution is of type IIB theory.
Again the application of $T_i$ generate a new solution of IIA theory. 
\begin{eqnarray}
 \label{T_j}
 {ds''}_{10}^2 &=& - e^{2\lambda'^{0}}dt^2 + \sum_{\mu \ne \{i,j,k\} } 
e^{2 \lambda'^\mu} (dx^\mu)^2 + e^{-2 \lambda'^j} (dx^j)^2
+e^{-2 \lambda'^i} (dx^i)^2 + e^{2 \lambda'} d r^2 
+ e^{2 \sigma'} d \Omega^2_{m, \epsilon}\nonumber\\
 \phi'' &=& \phi - \lambda'^i - \lambda'^j = 
\frac{1}{2} \lambda^k -\lambda^i - \lambda^j \;.
\end{eqnarray}
Dimensional upliftment to 11 dimensional theory can be done via
\begin{equation}
  \label{upk}
  ds'^2_{11} = e^{-\frac{2}{3} \phi''} \; {ds''}_{10}^2 
  + e^{\frac{4}{3} \phi''} \, (dx^{k})^2 \;.
\end{equation}
Using equation (\ref{T_j}) in (\ref{upk}) one finds
\begin{equation}
 d s'^2_{11} = - e^{2{\lambda''}^0}dt^2 + \sum_{\mu=1}^{q} 
e^{2 {\lambda''}{^\mu}} (dx^\mu)^2 + e^{2 \lambda''} d r^2 
+ e^{2 \sigma''} d \Omega^2_{m, \epsilon}\;,
\end{equation}
where these $\lambda{''^i}$'s are given in terms of $\lambda^i$'s by
(using equation (\ref{down_k})),
\begin{eqnarray}
 \lambda{''^i} &=& \lambda^j - \frac{2}{3} (\lambda^i+\lambda^j+\lambda^k)
  \nonumber \\
 \lambda{''^j} &=& \lambda^i - \frac{2}{3} (\lambda^i+\lambda^j+\lambda^k)
  \nonumber \\
 \lambda{''^k} &=& \lambda^k - \frac{2}{3} (\lambda^i+\lambda^j+\lambda^k)
  \nonumber \\
 \lambda{''^l} &=& \lambda^l +  \frac{1}{3} (\lambda^i+\lambda^j+\lambda^k) 
\;\;\;\;\;\;\;  \forall \; l \ne \{i,j,k\}
\end{eqnarray}
In general, simplifying notation, we can write,  
application of U duality $\uparrow_k T_i T_j \downarrow_k$ in 
(\ref{dsgen11}), transforms the scale factors, $\lambda^i$s, to
$\lambda'^i$s, given by
\begin{eqnarray}
\label{lambdareln}
& & \lambda'^i = \lambda^j - 2 \lambda \; \; , \; \; \;
\lambda'^j = \lambda^i - 2 \lambda \; \; , \; \; \; 
\lambda'^k = \lambda^k - 2 \lambda \nonumber \\
& & \lambda'^l = \lambda^l + \lambda \; \; , \; \; \;
l \ne \{i, j, k\}  \; \; , \; \; \;
\lambda = \frac{\lambda^i + \lambda^j + \lambda^k}{3} \; \; .
\label{uduality}
\end{eqnarray}

\subsection{Application of U duality relations in Black Holes} \label{sec:Ubh}
Note that, the U duality
relations follow as long as the directions involved in the U
duality operations are isometry directions. 
So the relations are valid for the geometry described by \ref{dsbh}.
We show now relations among $\lambda$'s 
follow from U duality relations. Consider a solution 
of $M2$ brane along $(x^1,x^2)$. Furthermore take, $(x^3,x^4,x^5)$ are compact
and isometry directions. An obvious symmetry implies 
\begin{equation}
 \label{obv2bh1}
 \lambda^1 = \lambda^2 \;,
\end{equation}
and
\begin{equation}
 \label{obv2bh2}
 \lambda^3 = \lambda^4 = \lambda^5 \;.
\end{equation}
Directions $x^a$ $(\in \{\Omega_4 \; \text{and} \; r\})$ are also transverse to 
brane directions. So we may assume 
\begin{equation}
 \label{obv2bh3}
 \lambda^3 = \lambda^4 = \lambda^5 =\lambda^6 = \lambda^7 = \lambda^8
= \lambda^9 = \lambda^{10} \;.
\end{equation}
Now apply U duality operations $\downarrow_5 T_3 T_4 \uparrow_5$. They transform
$M2$ brane to $M5$ brane.
\begin{displaymath}
 \xymatrix{
           M2 (12) \ar[r]^{\downarrow_5} & D2 (12) \ar[r]^{T_4} & D3 (124)
           \ar[r]^{T_3} & D4 (1234) \ar[r]^{\uparrow_5} & M5 (12345)
          } \; .
\end{displaymath}
This new metric of $M5$ branes may be given by
\begin{equation}
 d s'^2_{11} = - e^{2\lambda'^0}d t^2 + \sum_{i=1}^{10} 
e^{2 \lambda{'^i}} (d x^i)^2 + e^{2 \lambda'} d r^2 
+ e^{2 \sigma'} d \Omega^2_{m, \epsilon}\;.
\end{equation}
We can find $\lambda{'^i}$'s using equations (\ref{lambdareln}).
There are obvious symmetry relations for $M5$ brane, namely, 
\begin{equation}
 \label{obv5bh}
 \lambda'^1 = \lambda'^2 = \lambda'^3 = \lambda'^4 = \lambda'^5 \;,
\;\;\;
 \lambda'^6 = \lambda'^7 = \lambda'^8 = \lambda'^9 = \lambda'^{10} \;.
\end{equation}
So now one can write the relations among $\lambda$'s as 
\begin{equation}
 \label{25relnbh}
 \lambda^\parallel + 2 \lambda^\perp = 0 , \;\;\;
 2 \lambda'^\parallel + \lambda'^\perp =0 \;,
\end{equation}
where the superscripts $\parallel \;$ and $\perp \;$ denote
spatial dimensions parallel and transverse to the branes
respectively. 
Note that, to find these relations, we have used
duality relations only. Explicit form of $\lambda^\alpha$s can only be known
by solving equations of motion and putting proper boundary conditions, like
asymptotic flatness.

For the extremal $22'55'$ configuration $(12, 34, 13567,
24567) $, the transverse space is three dimensional and 
U duality relations come out, following above steps, to be
\begin{equation}\label{2255relnbh}
\lambda^1 + \lambda^4 + \lambda^5 = 
\lambda^2 + \lambda^3 + \lambda^5 = 0 \; \; . 
\end{equation}
Note that 
obvious symmetry relations for $22'55'$ black hole are
\begin{equation}\label{obv2255bh}
\lambda^5 = \lambda^6 = \lambda^7
\; \; , \; \; \; \lambda^8 = \lambda^9 = \lambda^{10} \; \; .
\end{equation}
One can verify from the explicit solution that these relations are true. 
See \cite{Bhowmick:2012dc} for detail.

We further illustrate the U duality method by interpreting a U
duality relation $\sum_i c_i \lambda^i = 0 $ as implying a
relation among the components of the energy-momentum tensor
$T_{A B} $. The relations thus obtained are indeed obeyed by
the components of $T_{A B}$ calculated explicitly.

Consider now the case of $2$ branes or 5 branes. We assume that
$P_a = P_\perp \;$ which is natural since $\theta^a \;$
directions are transverse to the branes. Applying the U duality
relations in equation (\ref{25relnbh}) then implies, for both 2
branes and 5 branes, the relation
\begin{equation}\label{zbh}
P_\parallel = P_0 + P_\perp + P_R \;,
\end{equation}
among the components of their energy-momentum tensor.
See equations (\ref{emt2}) and (\ref{emt5}).
Note that
it is also natural to take $P_0 = P_\parallel \;$ since $x^0 = t
\;$ is one of the worldvolume coordinates and may naturally be
taken to be on the same footing as the other ones $(x^1, \cdots,
x^p) \;$. Equation (\ref{zbh}) then implies that $P_\perp = - P_R
\;$. The relation between $P_\parallel \;$ and $P_R$ is to be
specified by an equation of state which
is given in equations (\ref{emt2}) and (\ref{emt5}).

\section{Exotic Branes} \label{sec:Exbh}
The formulation we showed in the previous section can also be applied for exotic branes. 
In this section, we show that explicitly.
In string theory/M-theory, exotic branes are always present. They are co-dimension 2
extended objects,
that is, in string theory they are 7-dimensional objects and in M-theory they are 
8-dimensional objects. 
In string theory, they are related to D-branes by 
S and T dualities. In other words, if we T or S-dualise D-brane of type IIA or IIB 
theory we may end up in exotic branes. For example take a $D5(12345)$
brane of type IIB theory stretched along $x^1,x^2,x^3,x^4,x^5$. Again as before 
our coordinates are $x^0,x^1,\cdots,x^9$; $x^0$ being timelike. Let us take, $x^6$ 
and $x^7$ are also compact. Now perform duality operations 
$ST_6T_7$. \footnote{Here
we follow a notation similar to \cite{Obers:1998fb}. That is $A^b_n$ means a $(A+b)$-brane, 
whose mass linearly depends on $A$ special dimensions, quadratically depends on 
$b$ special dimensions and subscript $n$ denotes branes mass is proportional to
$g_s^{-n}$. The first set 
of numbers in the bracket indicate $A$ spacelike worldvolume directions, the second set 
indicate where T-duality is performed, that is $b$ directions.}
\begin{displaymath}
 \xymatrix{
          D5(12345)~~ \ar[r]^{S} &  ~~NS5 (12345)~~ \ar[r]^{T_6} & 
          KKM5(12345,6) \ar[r]^{T_7} & ~~~5^2_2 (12345,67) 
          } \; .
\end{displaymath}
An S-duality on $D5$ brane generates an $NS5$ brane. $NS5$ branes are the source of an NS-NS 2-form 
field, $B_{\mu\nu}$. $NS5$ branes couple to $B_{\mu\nu}$ magnetically. So only
non-vanishing components of $B$'s are $B_{\mu\nu}$ with $\{\mu,\nu\}=\{6,7,8,9\}$.
In case of black branes, which are spherically symmetric in transverse space, in 
properly chosen gauge, only non-vanishing $B$ is $B_{67}$. Now a T-duality on $x^6$
generates Kaluza-Klein monopole. This T-duality also generates a cross-component
in metric $g_{\mu\nu}$, but $B$-field becomes zero. Another T-duality along $x^7$ 
generates exotic 5 brane $5^2_2$. Here again, cross component of metric is zero but
new metric may now depend on the angular coordinate so that it becomes a function
of $r$ and $\theta$. Equations of motion become more complicated. So it is not
obvious what will happen to the relations among various components of metric and
energy-momentum tensor.

\subsection{General Exotic Branes in M-Theory}
In M-theory $M2$ branes or $M5$ branes, while U dualise, may end up in exotic branes.
We will give an example later. In such cases, black exotic branes metric has to be function of
one angular direction. So
line element of general exotic black brane solution can be written as 
\begin{equation} \label{dsexotic}
 d s^2 = - e^{2 \lambda^0(r,\theta)} d t^2 
+ \sum_{i=1}^8 e^{2 \lambda^i(r,\theta)} (d x^i)^2 + e^{2 \lambda(r,\theta)} d r^2 
+ e^{2 \sigma(r,\theta)}  d \theta^2 \; .
\end{equation}
We take energy-momentum tensor as 
\begin{eqnarray}
 T^i{}_i &=& P_i ~~~~ \forall i=1,\cdots,8 \;, \nonumber \\
 T^r{}_r &=& P_R \;, \nonumber \\
 T^\theta{}_\theta &=& P_\Theta \;.
\end{eqnarray}
One can see, there is a non-zero $r-\theta$ component of Einstein tensor and
so $T_{r\theta}$ is also non-zero as well. We take 
\begin{equation}
 T^r{}_\theta = P_{R\Theta} \;.
\end{equation}
Note that scale factors are now functions of $r$ and $\theta$, so it is
natural to treat $r$ and $\theta$ directions separately. Therefore, we define
$\Lambda$ as $\Lambda = \sum_{i=0}^8 \lambda^i$.
Equations of motion are very similar to equations (\ref{t21bh}), (\ref{t22bh})
and (\ref{t23bh}) with obvious differences.
Equations of motion for the ansatz (\ref{dsexotic}) turn out to be
\begin{eqnarray}
\label{exoeomrr}
&e^{-2 \lambda } \left[\Lambda _r \sigma _r+
\frac{1}{2} \left(\Lambda _r^2-\sum_i {\lambda^i_r}^2\right)\right]
+
e^{-2 \sigma } \left[\Lambda _{\theta \theta} -\Lambda _{\theta } \sigma _{\theta } 
+\frac{1}{2}\left(\Lambda _\theta^2+\sum_i {\lambda^i_\theta}^2\right)\right]
= P_R \;,&\\
\label{exoeomthth}
&e^{-2 \lambda }\left[\Lambda _{rr}-\Lambda _r \lambda_r+
\frac{1}{2} \left(\Lambda _r^2+\sum_i {\lambda^i_r}^2\right)\right]
+
e^{-2 \sigma } \left[\Lambda _{\theta } \lambda_{\theta }+
\frac{1}{2}\left(\Lambda _\theta^2-\sum_i {\lambda^i_\theta}^2\right)\right]
= P_\Theta \;, & \\
\label{exoeomii}
&e^{-2 \lambda }\left[\lambda^i_{rr}+(\Lambda _r+\sigma_r -\lambda_r) \lambda^i_r\right]
+ 
e^{-2 \sigma }\left[\lambda^i_{\theta\theta}+(\Lambda _\theta-\sigma_\theta +\lambda_\theta) \lambda^i_\theta\right]
= -P_i + \frac{1}{9}(\sum_i P_i + P_R+P_\Theta) \;.&
\end{eqnarray}
There are two more equations one can write, one for $r\theta$ component and one 
conservation equation. They are not independent. Solving these equations, with proper 
boundary conditions, enables one to get exotic black branes.

\subsection{Exotic Branes of M-Theory}
We illustrate this
idea in M-theory by one explicit example, black brane solution for $5^3$. An 
explicit solution may be found in \cite{deBoer:2012ma}. 
\begin{equation}
 \label{5^3soln}
 ds^2 = H^{-1/3}W^{2/3}\left(-dt^2+\sum_{i=1}^5 (dx^i)^2\right) + 
        H^{2/3}W^{-4/3}\sum_{i=6}^8 (dx^i)^2 + 
         H^{2/3}W^{2/3}\left(dr^2+r^2 d\theta^2\right) \;,
\end{equation}
where $H = h + b \ln \frac{\mu}{r}$ and $W^2 = H^2+b^2\theta^2$. Here $b$ is the 
charge, and given in terms of number of branes and radii of $x^6$, $x^7$ and 
$x^8$ circle as $b = \frac{N R_6 R_7 R_8}{2\pi l_p^3}$. 
$\mu$ is a cut off in energy scale which has to be present in any co-dimension 2
brane solutions.
3-form field is
given by
\begin{equation*}
 C_{678} = \frac{b \theta}{W^2}. 
\end{equation*}
Note that field strength has now two components $F_{678\theta}$ and 
$F_{678r}$. So there is a non-zero $r\theta$-component of energy
momentum tensor. Note the non-geometric nature of the above metric.
Metric and fields depend on $\theta$ in such a manner that they do not
come back to the initial values when angle goes from $\theta$ to $\theta + 2\pi$. 
That is metric 
and fields are not periodic functions of $\theta$.

We see from explicit calculation that
\begin{equation}
 \label{exoemt1}
 T^0{}_0 = T^\parallel{}_\parallel = - T^\perp{}_\perp = \frac{b^2}{4 r^2 H^{8/3} W^{2/3} }\;. 
\end{equation}
Here $\parallel$ and $\perp$ indicate $(x^1,\cdots,x^5)$ and $(x^6,x^7,x^8)$ 
respectively, though strictly speaking $(x^6,x^7,x^8 )$-directions are not transverse to the branes.
This is an 8-brane in M-theory.
One also finds from explicit calculation that,
\begin{equation}
 \label{exoemt2}
 T^r{}_r = - T^\theta{}_\theta = -\frac{b^2 \left(b^4 \theta ^4-6 b^2 \theta ^2 H^2+H^4\right)}
 {4 r^2 H^{8/3} W^{14/3} } \;.
\end{equation}
Now one needs equations of state which relate $T^r{}_r$ and $T^r{}_\theta$
to $T^\perp{}_\perp$. It turns out these equations of state are dependent
on $r$ and $\theta$ by a highly non-linear functions of them. These 
functions are monotonic functions of $\theta$, which is characteristic of
non-geometry. 

Equations (\ref{exoemt1}), (\ref{exoemt2}) and equation of motion (\ref{exoeomii})
imply constraining relations among scale factor.
\begin{eqnarray}
 \lambda^0 &=& \lambda^\parallel \nonumber \\
 2\lambda^\parallel + \lambda^\perp &=& 0 \;.
\end{eqnarray}
One can see that above relations are same as in $M5$ brane case. In the next section we 
show that same relations come out from U-duality.

\subsection{U duality Relations In Exotic Branes}
We consider two examples, $5^3$ and $2^6$ to illustrate the idea. 

\subsubsection*{$5^3$}
Let us start with M5-branes, and apply U-duality operations of the form 
$\downarrow_1 T_6 S T_7 T_8 S T_6 \uparrow_1$. They transform M5 brane into one 
of the exotic brane of M-theory. 
\begin{displaymath}
 \xymatrix{
           M5 (12345) \ar[r]^{\downarrow_1} & D4 (2345) \ar[r]^{T_6} & D5 (23456)
           \ar[r]^{S~} & NS5 (23456) \ar[d]^{T_7 T_8} \\
           5^3 (12345,678)  & 
           4^3_3 (2345,678)\ar[l]_{~~\uparrow_1}  & 5^2_3 (23456,78) 
           \ar[l]_{~~T_6}  & 5^2_2 (23456,78) \ar[l]_{~S}
          } \; 
\end{displaymath}
General supergravity solution of black M-brane is given by equation 
(\ref{dsgen11}). In case of black M5 brane, we have obvious symmetries
\begin{eqnarray*}
 &&\lambda^1 =  \lambda^2 = \lambda^3 = \lambda^4 = \lambda^5 = \lambda^\parallel\\
 &&\lambda^6 =  \lambda^7 = \lambda^8 = \lambda^9 = \lambda^{10} = \lambda^\perp \;.
\end{eqnarray*}
Just like before (see equation (\ref{down_k})) doing a dimensional reduction along
$x^1$ we find 
\begin{eqnarray}
 \label{down1}
 \phi &=& \frac{3}{2} \lambda^{1}  \nonumber \\
 \lambda''^\mu &=& \lambda^\mu + \frac{1}{3} \phi
 = \lambda^\mu +\frac{1}{2} \lambda^1 \;,
\end{eqnarray}
where $\mu = 0,2,3,\cdots,10$. Then $T_6 S$ transform dilaton and metric into
\begin{eqnarray} \label{T_6S}
 \phi &=& \lambda^6 - \lambda^1 \nonumber \\
 \lambda'''^\mu &=& \left\{\lambda^0+\frac{\lambda^6}{2},
 \lambda^2+\frac{\lambda^6}{2},\lambda^3+\frac{\lambda^6}{2},
 \lambda^4+\frac{\lambda^6}{2},\lambda^5+\frac{\lambda^6}{2}, \right. \nonumber \\ 
 &&\left. -\lambda^1-\frac{\lambda^6}{2}, \frac{\lambda^6}{2}+\lambda^7,
   \frac{\lambda^6}{2}+\lambda^8,
   \frac{\lambda^6}{2}+\lambda^9,\frac{\lambda^6}{2}+\lambda^{10}\right\} .
\end{eqnarray}
Last S-duality also generates an NS-NS 2-form field. This 2-form field has to be 
spherically symmetric in transverse space $(x^9,x^{10})$. We write this transverse
space metric as $e^{2\lambda^9}dr^2+e^{\lambda^{10}}r^2 d\theta^2$. Then in proper
gauge 2-form field may have only $B_{78}$ component, and can be taken as 
$B_{78} = b(r) \theta$.\footnote{In general for spherically symmetric case 
$B_{78} = \epsilon_{02345678r\mu} b(r) x^\mu$. So field strength is only 
$r$-dependent and has component $F_{78\theta}$. However one T-duality 
converts $B_{78}$ into metric components
$g_{78}$ and one more T-duality converts $g_{78}$ to $B'_{78}$
which is now function of both $r$ and $\theta$. So field strength has components
$F_{78\theta}$ and $F_{78r}$. They are also function of $r$ and $\theta$,
so that non-geometry is introduced.} 
In this black exotic brane case, $b(r)$ is constant.
Now application of rest of the duality operations 
$T_7 T_8 S T_6 \uparrow_5$ uplift the metric to 11-dimension again. 
This new 11 dimensional
line element looks like
\begin{equation}
 d s'^2_{11} = - e^{2{\lambda'}^0}dt^2 + \sum_{\mu=1}^{10} 
e^{2 {\lambda'}{^\mu}} (dx^\mu)^2 \;.
\end{equation}
Here $\lambda'^\mu$'s are given by
\begin{eqnarray} \label{lambda5^2_2}
  \lambda'^\mu = &&\left\{\lambda _0+\frac{\ln (L)}{6},
  \lambda^1+\frac{\ln (L)}{6},\lambda^2+\frac{\ln (L)}{6},
  \lambda^3+\frac{\ln (L)}{6},
  \lambda^4+\frac{\ln (L)}{6},\lambda^5+\frac{\ln (L)}{6}, \right. \nonumber \\
   &&\left.\lambda^6-\frac{\ln (L)}{3},
   \lambda^8-\frac{\ln (L)}{3},\lambda^7-\frac{\ln
   (L)}{3},\lambda^9+\frac{\ln (L)}{6},
   \lambda^{10}+\frac{\ln (L)}{6}\right\} \;,
\end{eqnarray}
where $L=\left(b^2 \theta ^2+e^{2 \left(\lambda^6+
\lambda^7+\lambda^8\right)}\right)$. One can now use 
obvious symmetries of M5 brane and the equation (\ref{25relnbh}) 
to get a relation among
various $\lambda'$s for the case of exotic 8-brane, $5^3$.
One can see immediately the symmetries of $5^3$ from expression of 
$\lambda'^\mu$ of equation \ref{lambda5^2_2}
\begin{eqnarray}
  &&\lambda'^1 =  \lambda'^2 = \lambda'^3 = \lambda'^4 = \lambda'^5 = 
  \frac{1}{6} \ln \left(b^2 \theta ^2 e^{6 \lambda^\parallel }+e^{-6 \lambda^\parallel }\right) = 
  \lambda'^\parallel \;,\;\; \mbox{(say)}, \nonumber \\
 &&\lambda'^6 =  \lambda'^7 = \lambda'^8 = 
 -\frac{1}{3} \ln \left(b^2 \theta ^2 e^{6 \lambda^\parallel }+e^{-6 \lambda^\parallel }\right) = 
 \lambda'^\perp \;,\;\; \mbox{(say)}, \nonumber \\
  &&\lambda'^9 = \lambda'^{10} = 
  \frac{1}{6} \ln \left(b^2 \theta ^2 e^{3\lambda^\parallel}+e^{-9 \lambda^\parallel }\right) =
  \lambda'^\odot \;,\;\; \mbox{(say)}\;.
\end{eqnarray}
Now we can see U-duality implies a relation between $\lambda'^\parallel$ and $\lambda'^\perp$.
\begin{eqnarray}
 \lambda'^\perp + 2 \lambda'^\parallel &=& 0 \;.
\end{eqnarray}

\subsubsection*{$2^6$}
Another exotic brane in M-theory is $2^6$. 
One obtains a $2^6~(12,345678)$
brane by applying duality operations,
$\downarrow_1 T_6 S T_7 T_8 S T_6 T_5 T_4 T_3 S \uparrow_1$ on an $M5~(12345)$ brane.
As in the case of $5^3$ brane we start black $M5$-brane metric as in equation
\ref{dsgen11} and metric along $(x^9,x^{10})$ as 
$e^{2\lambda^9}dr^2+e^{\lambda^{10}}r^2 d\theta^2$. Application of said duality
transformations on $M5$ brane metric we get
\begin{equation}
 d s'^2_{11} = - e^{2{\lambda'}^0}dt^2 + \sum_{\mu=1}^{10} 
e^{2 {\lambda'}{^\mu}} (dx^\mu)^2 \;,
\end{equation}
with $\lambda'^\mu$'s are given by
\begin{eqnarray}
  &&\lambda'^1 =  \lambda'^2 = 
  \frac{1}{3} \ln \left(b^2 \theta ^2 e^{6 \lambda^\parallel }+e^{-6 \lambda^\parallel }\right) = 
  \lambda'^\parallel \;,\;\; \mbox{(say)}, \nonumber \\
 &&\lambda'^3 = \lambda'^4 = \lambda'^5 = \lambda'^6 =  \lambda'^7 = \lambda'^8 = 
-\frac{1}{6} \ln \left(b^2 \theta ^2 e^{6 \lambda^\parallel }+e^{-6 \lambda^\parallel }\right) = 
 \lambda'^\perp \;,\;\; \mbox{(say)}, \nonumber \\
  &&\lambda'^9 = \lambda'^{10} = 
  \frac{1}{3}\ln \left(b^2 \theta ^2 e^{-3 \lambda^\parallel}+e^{-15 \lambda^\parallel}\right) =
  \lambda'^\odot \;,\;\; \mbox{(say)}\;.
\end{eqnarray}
In deriving above equations, we use again obvious symmetry of $M5$ branes and equation
(\ref{25relnbh}). 
The relation between $\lambda'^\parallel$ and $\lambda'^\perp$ turns out
to be
\begin{eqnarray}
 2\lambda'^\perp + \lambda'^\parallel =0 \;.
\end{eqnarray}

Using equation of motion (\ref{exoeomii}) and the above relation among $\lambda$'s 
one finds a relation similar to (\ref{zbh}),
\begin{equation}
 \label{zexo}
 P_\parallel = P_0+P_\Theta+P_R \;.
\end{equation}
Equation (\ref{zexo}) is on of the main results of this paper. A possible application of 
the equation is discussed in section \ref{sec:TD}.

\section{Black Branes in String Theory and Dualities} \label{sec:Stringbh}

\subsection{General Black Branes in String Theory}
In this subsection, we see that a similar relation exists for the black branes
in the string theory. In the next subsection, we will show, 
just like M-theory case, for 
certain supergravity solutions S and T-duality can be used to get relations
among various metric components and dilaton, and hence relations among
energy-momentum tensor. To illustrate this, we consider 
a general solution of string theory (type II supergravity).
Line element pf $p$-branes, $ds_{Dp}$ in Einstein frame is 
\begin{equation}\label{dsStrp}
ds_{Dp}^2 = - e^{2\lambda^0_p} Z d t^2 + 
\sum_{i=1}^{p} e^{2 \lambda^\parallel_p} (d x^i)^2+
\sum_{i=p+1}^{q} e^{2 \lambda^\perp_p} (d x^i)^2
+e^{2\lambda} \frac{dr^2}{Z} + e^{2\sigma} d\Omega_{m,\epsilon}^2 \;,
\end{equation}
and dilaton, $\phi_p\,=\,\phi_p(t,r)$. Here we assume $(x^1, \cdots, x^q)$ are
compact and are isometry directions.
Here $i=1,\cdots,p$ are directions parallel to Dp-branes. Obvious symmetries
ensure us to take all $\lambda^i$s parallel to branes as equal and we denote 
them as before by $\lambda^\parallel_p$, similarly for $\lambda^\perp_p$. 
Here all $\lambda$s, $Z$ and $\sigma$ are functions of $r$.
$d\Omega_{m,\epsilon}$ is metric of constant curvature 
$m$-dimensional space. 

Bosonic part of the supergravity action for type II string theory in Einstein frame is
\begin{equation}
\label{10dAction}
 S =  \frac{1}{16 \pi G_{10}} \int d^{10}x \, \sqrt{-g} 
 \left( R - \frac{1}{2} (\partial \phi)^2
   - \sum_p \frac{e^{a_p\phi}}{2 (p+2)!}  F_{p+2}^2 \right)\; ,
\end{equation}
where $F_{p+2}=dA_{p+1}$ is $p+2$-form field strength coupled to $p$-brane. $T_{AB}$
is given by
\begin{equation}
 \label{bhstrTAB}
 T_{AB} =  \frac{1}{2} \partial_A \phi \partial_B \phi - \frac{1}{4} g_{AB} (\partial \phi)^2
          +\sum_p \frac{e^{a_p\phi}}{2 (p+2)!} \; \left[(p+2) F_{AM_1\cdots M_{p+1}} 
          F_B{}^{M_1\cdots M_{p+1}}
          -\frac{1}{2} g_{AB} \, F_{p+2}^2 \right] \; .
\end{equation}
Here $a_p$ is a $p$ dependent factor, its value also depends on the type of brane. For
$Dp$ brane, $a_p = \frac{3-p}{2}$.
We also have another ``component'', $T_\phi$ coming from $\phi$-variation of action
\begin{equation}
 \delta_\phi S = -\int d^{10}x \; \sqrt{-g} \; (\triangledown^2\phi+T_\phi) \;\delta\phi\; ,
\end{equation}
where $\delta_\phi$ denote variation with respect to $\phi$.
Equations of motion are now Einstein equations together with
\begin{equation}\label{Tfhi}
 \nabla^2 \phi = -T_\phi \;.
\end{equation}

Again non-vanishing components of $T^A{}_B$
are given by $T^r{}_r = P_R$ and $T^\alpha{}_\alpha
= P_\alpha$, where $\alpha = (0, i, a)$.
These components can be written explicitly as
\begin{eqnarray*}
 P_0 = P_\parallel &=& - \frac{Z}{4} e^{-2\lambda} \phi_r^2 + \frac{e^{a_p\phi}}{4} F_{01\cdots p r} F^{01\cdots p r} \\
 P_\perp = P_a = -P_R &=&  - \frac{Z}{4} e^{-2\lambda} \phi_r^2 - \frac{e^{a_p\phi}}{4} F_{01\cdots p r} F^{01\cdots p r} \;.
\end{eqnarray*}
Similarly $T_\phi$ is given by
\begin{equation*}
 T_\phi = \frac{a_p}{2} e^{a_p\phi} F_{01\cdots p r} F^{01\cdots p r} \;.
\end{equation*}

The equations of motion then may be written as
\begin{eqnarray}
\Lambda_r^2 - \sum_\alpha (\lambda^\alpha_r)^2 & = & 
2 \; P_R + \epsilon \; m (m - 1) e^{- 2 \sigma} \;, \label{t21bhstring} \\
Ze^{-2\lambda}\left[\lambda^\alpha_{r r} + (\Lambda_r-\lambda_r) \lambda^\alpha_r\right] & = & 
- \; P_\alpha + \frac{1}{8} \; (P_R + \sum_\beta P_\beta) 
\nonumber \\
& & + \; \epsilon \; (m - 1) e^{- 2 \sigma} \; \delta^{\alpha a}
\label{t22bhstring} \;,\\
Ze^{-2\lambda}\left[\phi_{r r} + (\Lambda_r-\lambda_r) \phi_r\right] &=& -T_\phi 
\label{tphi} \;, \\
{P_{R}}_r + P_R \Lambda_r - \sum_\alpha P_\alpha \lambda^\alpha_r & = & 0 \;,
\label{t23bhstring} 
\end{eqnarray}
where $\Lambda = \sum_\alpha \lambda^\alpha = \lambda^0 + \sum_i
\lambda^i + m \sigma \;$ and the subscripts $r$ denote
$r$-derivatives. In case of intersecting branes $P_\alpha = \sum_I P_{\alpha\,(I)}$, 
$P_R = \sum_I P_{R(I)}$ and equation (\ref{t23bhstring}) may be written as
\begin{equation}
 \label{t23bhI}
 {P_{R(I)}}_r + P_{R(I)} \Lambda_r - \sum_\alpha p_{\alpha\,(I)} 
\lambda^\alpha_r = 0\;.
\end{equation}
This is because of, as we already claimed, $T^A{}_{B\,(I)}$ obey conservation
equation separately.

If we assume $P_{\alpha \,(I)} = - (1 - u^I_\alpha) \;P_{R(I)} $, then equation
(\ref{t23bhI}) can be solved. The solution is found to be
\begin{equation}
\label{f_I}
P_{R(I)} = - \; e^{l^I - 2 \Lambda} \; \; , \; \; \; 
l^I = \sum_\alpha u^I_\alpha \lambda^\alpha +l^I_0 \;.
\end{equation}
Now we define the matrices $G_{\alpha \beta}$ and ${\cal G}^{I J}$ as
\begin{equation}
\label{defnG}
 G_{\alpha \beta} = 1 - \delta_{\alpha \beta} 
\; \; \; , \; \; \; \;
{\cal G}^{I J} = \sum_{\alpha, \beta} 
G^{\alpha \beta} \; u^I_\alpha \; u^J_\beta \;  ,
\end{equation}
where $G^{\alpha\beta}$ is the inverse of $G_{\alpha\beta}$ and is given by
\begin{equation}
 G^{\alpha\beta} = \frac{1}{8} -\delta^{\alpha\beta} \;.
\end{equation}
If we define a 
new coordinate $\tau$ by $d\tau = e^{- \Lambda+\lambda} \; dr$, equation (\ref{t22bhstring})
becomes 
\begin{eqnarray}
\label{lambdatautau}
e^{-2\Lambda} \; \lambda^\alpha_{\tau \tau} =  
- \; P_\alpha + \frac{1}{8} \; (P_R + \sum_\beta P_\beta)
+ \; \epsilon \; (m - 1) e^{- 2 \sigma} \; \delta^{\alpha a} \;.
\label{t22bhtau} 
\end{eqnarray}
If one multiplies both sides by $u_\alpha^I$ and take a sum over $\alpha$ and then 
use equation (\ref{f_I}) in (\ref{t22bhtau}) one finds
\begin{equation}
\label{genlI}
l^I_{\tau \tau} = - \; \sum_J {\cal G}^{I J} \; e^{l^J} 
+ \sum_{a \in \Omega} u_a^I \; \epsilon \; (m - 1) \; e^{2 (\Lambda - \sigma)}
\; \; .
\end{equation}
Specific values of ${\cal G}^{I J}$ depend on 
intersecting configuration. In case 
of black holes we can calculate them from the explicit expression of
energy-momentum tensor or by using duality. In section {\bf \ref{sec:stringd}} we
discuss duality method.

If the components of energy-momentum tensor follow a relation like
\begin{equation}
 \label{rlnbh}
 \sum_\alpha c_\alpha 
\left( -P_\alpha + \frac{1}{8} \; (P_R + \sum_\beta P_\beta) \right) = 0
\end{equation}
then this immediately implies a relation among metric components $\lambda^\alpha$
and $\sigma$ from equation \ref{t22bhstring} or \ref{lambdatautau}. We will see, 
for various black brane solutions,
explicit calculation and using duality technique give us same relations.
For example, when $\alpha \ne a$, then equations (\ref{lambdatautau}) and
(\ref{rlnbh}) imply 
\begin{equation}
 \label{weakbh}
 \sum_\alpha c'_\alpha \lambda^\alpha =0 \;.
\end{equation}
We will see some examples in the next section, which show, 
relation like \ref{rlnbh} does follow.
Moreover, explicit computation shows there is a linear relation among $P$'s and $T_\phi$. They 
in tern imply
\begin{equation}
 \label{weakbhphi}
 \sum_\alpha \tilde{c}_\alpha \lambda^\alpha + \tilde{c}_\phi \phi = 0 \;.
\end{equation}
Any particular intersecting configuration gives specific values of 
$c_\alpha$, $c'_\alpha$, $\tilde{c}_\alpha$ and $\tilde{c}_\phi$.

\subsubsection*{Example: Extremal $D1$ Branes} 
Let us consider a stack of $D1$ branes wrapped around $x^1$ direction, also take 
$(x^2, \cdots, x^5)$ are compact. The solution is well known. It can be written as
\begin{equation} \label{D1}
ds^2 = - e^{2\lambda^0}  d t^2 + 
e^{2 \lambda^\parallel} (d x^1)^2+
\sum_{i=2}^{5} e^{2 \lambda^\perp} (d x^i)^2
+e^{2\lambda} {dr^2} + e^{2\sigma} d\Omega_{3}^2 
\end{equation}
Dilaton is given by $\phi(r)$
and non-vanishing components of RR form field are $C_{01}(r)$. Explicit calculation
shows that 
\begin{equation}
 \label{emtD1}
 T^0{}_0 = T^\parallel{}_\parallel =  
 \frac{5}{3} T^r{}_r 
 =-\frac{5}{3} T^\perp{}_\perp =-\frac{5}{3} T^a{}_a \;.
\end{equation}
Now equation \ref{emtD1} and equation \ref{t22bhstring} imply 
\begin{equation}
 \lambda^\parallel + 3 \lambda^\perp =0 \;.
\end{equation}
Also from explicit calculation one can see 
\begin{equation}
 \phi - 4 \lambda^\perp = 0 \;.
\end{equation}

\subsubsection*{Example: Extremal $NS5$ Branes} 
Let us take another example of extremal black $NS5$-brane solution. These branes are 
wrapped around $(x^1,\cdots,x^5)$. Consider $x^6$ is also compact. The 
line element is given by
\begin{equation} \label{NS5}
ds^2 = - e^{2\lambda^0}  d t^2 + 
\sum_{i=1}^{5} e^{2 \lambda^\parallel} (d x^i)^2+
 e^{2 \lambda^\perp} (d x^6)^2
+e^{2\lambda} {dr^2} + e^{2\sigma} d\Omega_{2}^2 \; .
\end{equation}
Again explicit calculation shows that 
\begin{equation}
 T^0{}_0 = T^\parallel{}_\parallel =  
 \frac{5}{3} T^r{}_r 
 =-\frac{5}{3} T^\perp{}_\perp 
 =-\frac{5}{3} T^a{}_a \;.
\end{equation}
Above equations and explicit calculation with dilaton imply
\begin{eqnarray}
 3\lambda^\parallel +  \lambda^\perp &=& 0  \\
 3\phi - 4 \lambda^\perp &=& 0 \;.
\end{eqnarray}

\subsubsection*{Example: BPS Intersection of two sets of $D3$ branes }
Let us take two sets of $D3$ branes along $(x^1,x^2,x^3)$ and $(x^1,x^4,x^5)$. This
configuration is BPS configuration. We denote these two sets of $D3$ branes as $D3a$ and
$D3b$, and their charges as $h_a$ and $h_b$. Let us also take $x^6$ direction as compact.
The black hole solution for this configuration is given by
\begin{eqnarray}
ds^2 = - e^{2\lambda^0_p}  d t^2 + 
\sum_{i=1}^{5} e^{2 \lambda^i} (d x^i)^2
+e^{2\lambda} {dr^2} + e^{2\sigma} d\Omega_{3}^2 \;,
\end{eqnarray}
Again it is clear from detail of the solution 
\begin{equation}
 T^A{}_B = \sum_I T^A{}_{B(I)} \;.
\end{equation}
where $I$ indicates $D3a$ and $D3b$ branes. 
Conservation equations hold separately for each $I$. The 
relation among metric components becomes now
\begin{equation}
 \lambda^1+\lambda^4 = 0 \;,
\end{equation}
together with obvious symmetries
\begin{eqnarray}
 \lambda^1 &=& \lambda^2 \nonumber \\
 \lambda^3 &=& \lambda^4 \;.
\end{eqnarray}

\subsection{S and T-Duality Relations} \label{sec:stringd}

Here we show, the above mentioned relations follow from S and T-dualities of string theory.
To illustrate this, we consider an extremal black $p$-brane
solution. It is of the form given in equation (\ref{dsStrp}).
The subscript $Dp$ indicate metric is for $Dp$-branes.
This metric is of general black $p$-brane solution. $q$ is the total
number of compact directions, and $q+m=8$. This system physically
describes geometry created by $D$-brane localised in space. 
In the following analysis $\lambda$'s can be function of $r$ as well as
$t$. So the relations we find here can be used for black holes as well as
cosmological solutions.\footnote{A cosmological model similar to 
\cite{Kalyana Rama:2007ft} -- \cite{Bhowmick:2012dc} can be made in the 
framework of string theory.} 

This solution is in Einstein frame. To apply duality rules we first convert it
in string frame. String frame metric is
\begin{eqnarray}
\label{dsStrpS}
ds_{s,Dp}^2 &=& - e^{2\lambda^0_p + \frac{\phi_p}{2}} Z d t^2 + 
\sum_{i=1}^{p} e^{2 \lambda^\parallel_p + \frac{\phi_p}{2}} (d x^i)^2+
\sum_{i=p+1}^{q} e^{2 \lambda^\perp_p + \frac{\phi_p}{2}} (d x^i)^2 \nonumber \\
&+& e^{2\lambda + \frac{\phi_p}{2}} \frac{dr^2}{Z} + 
e^{2\sigma + \frac{\phi_p}{2}} d\Omega_{m,\epsilon}^2  \;.
\end{eqnarray}
Now if we perform a T-duality along $p^{th}$ direction we get $D\,(p-1)$
branes solution. 
\begin{eqnarray}
\label{dsStrpST}
ds_{s,Dp-1}^2 &=& - e^{2\lambda^0_p + \frac{\phi_p}{2}} Z d t^2 + 
\sum_{i=1}^{p-1} e^{2 \lambda^\parallel_p + \frac{\phi_p}{2}} (d x^i)^2+
e^{-(2 \lambda^\parallel_p + \frac{\phi_p}{2})} (d x^p)^2 \nonumber \\
&+& \sum_{i=p+1}^{q} e^{2 \lambda^\perp_p + \frac{\phi_p}{2}} (d x^i)^2 
+  e^{2\lambda + \frac{\phi_p}{2}} \frac{dr^2}{Z} + 
e^{2\sigma + \frac{\phi_p}{2}} d\Omega_{m,\epsilon}^2 \\
\label{phiT}
\phi_{p-1} &=& \phi_p -\frac{1}{2} 
ln (e^{2 \lambda^\parallel_p + \frac{\phi_p}{2}}) = \frac{3 \phi_p}{4} - 
\lambda^\parallel_p \;.
\end{eqnarray}

The Einstein frame metric for $D(p-1)$ solution can be found by multiplying 
$e^{-\phi_{p-1}/2}$ to $ds_{s,Dp-1}^2$.
\begin{eqnarray}\label{dsStrp-1E}
 ds_{Dp-1}^2 &=& e^{-\phi_{p-1}/2} ds_{s,Dp-1}^2 \nonumber \\
	     &= & - e^{2\lambda^0_{p-1}}Z d t^2 +
\sum_{i=1}^{p-1} e^{\frac{5 \lambda^\parallel_p}{2} + 
\frac{\phi_p}{8}} (d x^i)^2 + 
e^{-\frac{3 \lambda^\parallel_p}{2} - \frac{7\phi_p}{8})} (d x^p)^2
\nonumber \\
&+& \sum_{i=p+1}^{q} e^{2 \lambda^\perp_p + 
\frac{\lambda^\parallel_p}{2}\frac{\phi_p}{8}} (d x^i)^2   +ds_\perp^2 \;,
\end{eqnarray}
where $ds_\perp^2$ is metric for transverse space. 
Equations (\ref{dsStrpST}) and (\ref{phiT} have been used to get above 
expression.
This line element can be written as 
\begin{equation}\label{dsStrp-1}
ds_{Dp-1}^2 = -e^{2\lambda^0} Z dt^2 + 
\sum_{i=1}^{p-1} e^{2 \lambda^\parallel_{p-1}}
(d x^i)^2+ \sum_{i=p}^{q} e^{2 \lambda^\perp_{p-1}} (d x^i)^2
+ds_\perp^2 \;,
\end{equation}
where 
$\lambda_p$ can be given in terms of $\lambda_{p-1}$ by
 \begin{eqnarray}
 \label{lambdareln1}
  2\lambda^\parallel_{p-1} &=& \frac{5 \lambda^\parallel_p}{2}+\; 
\frac{\phi_p}{8}  \\
 \label{lambdareln2}
  2\lambda^\perp_{p-1} &=&
  -\frac{3}{2} \lambda^\parallel_p -\frac{7}{8} \phi_p =
  2\lambda^\perp_{p}+
 \frac{\lambda^\parallel_{p}}{2}+\frac{\phi_p}{8} 
 \end{eqnarray}
 and 
 \begin{eqnarray}\label{phireln}
  \phi_{p-1}=\frac{3 \phi_p}{2}-\lambda^\parallel_p \;.
 \end{eqnarray}
Simplification of equation (\ref{lambdareln2}) shows that,
\begin{equation}
 \label{strreln}
 \lambda_p^\perp \;+\; \lambda_p^\parallel \;+\; \frac{\phi_p}{2} \;=\;0\;.
\end{equation}
Now consider a $D3$ brane, S-duality of $D3$ brane gives $D3$ brane
and so $\phi_3=-\phi_3$, which implies $\phi_3=0$.
Consider $D2$ brane solution now, $\phi_2 = -\lambda^\parallel_3$, (using
equation (\ref{phireln})). Equation (\ref{strreln}) gives 
$\lambda^\parallel_3 = - \lambda^\perp_3$. Denoting $\lambda^\perp_3$ simply by
$\lambda^\perp$, one finds
\begin{equation}
 \lambda_3^\perp=\lambda^\perp \;,\;\;\, 
\lambda_3^\parallel=-\lambda^\perp \;,\;\;\;
\phi_3 = 0 \times \lambda^\perp \;.
\end{equation}
In general using induction, it is easy to show that,
\begin{equation}
 \lambda_p^\perp=\frac{p+1}{4}\lambda^\perp \;,\;\;\, 
\lambda_p^\parallel=-\frac{7-p}{4}\lambda^\perp \;,\;\;\;
\phi_p = (3-p) \lambda^\perp \;,
\end{equation}
where $\lambda$ is now the only parameter determining full line element.
If one uses different set of duality operations one can also find similar
relations for $F1$-string or $NS5$-brane. In general
\begin{equation}
\label{strrln}
\lambda_p^\perp=\frac{p+1}{4}\lambda^\perp \;,\;\;\, 
\lambda_p^\parallel=-\frac{7-p}{4}\lambda^\perp \;,\;\;\;
\phi_p = z(3-p) \lambda^\perp \;,
\end{equation}
where $z=1$ for $Dp$-brane and $z=-1$ for $F1$-string or $NS 5$-brane.

\subsubsection*{Intersecting Branes}
We use here above procedure for intersecting branes system. The procedure is
generally true for any mutually BPS intersecting branes configuration.
We show this with an example of $D3$-$D3$ system for illustration purpose. Two mutually BPS
$D3$ branes may intersect in one direction. 
So let us take first and second stack of $D3$ branes are wrapped
around $(x^1,x^2,x^3)$ and $(x^1,x^4,x^5)$ respectively.
In general we can write this black brane as
\begin{equation}
 \label{D3D3}
 ds^2 = -e^{2\lambda^0} dt^2 + \sum_{i=1}^9 e^{2\lambda^i} (dx^i)^2 \;.
\end{equation}
Here we have obvious symmetries
\begin{eqnarray} \label{obvD3D3}
 &\lambda^2 = \lambda^3 = \lambda^{D3a}& \text{(say)} \nonumber \\
 &\lambda^4 = \lambda^5 = \lambda^{D3b}& \text{(say)} \;.
\end{eqnarray}
One S-duality transforms dilaton $\phi \to -\phi$ but $D3$-$D3$ black hole
remains same. This implies $\phi=0$. Now apply two T-duality operations, 
$T_2T_3$. This is a $D1$-$D5$ black hole with metric 
\begin{equation}
 \label{D1D5}
 ds^2 = -e^{2\lambda^0} dt^2 +e^{2\lambda^1} (dx^1)^2+ \sum_{i=2,3}e^{-2\lambda^i} (dx^i)^2+\sum_{i=4}^9 e^{2\lambda^i} (dx^i)^2 \;,
\end{equation}
and dilaton
\begin{equation}
 \phi = -\lambda _2-\lambda _3 \;.
\end{equation}
Writing above solution in Einstein frame and using obvious symmetry \ref{obvD3D3}  we find
\begin{equation}
 \label{D1D5e}
 ds^2 = -e^{2\lambda'^0} dt^2 + \sum_{i=1}^9 e^{2\lambda'^i} (dx^i)^2 \;,
\end{equation}
where $\lambda'$'s are given by
\begin{eqnarray}
   \lambda'^\mu &=& \left\{\frac{\lambda ^{{D3a}}}{2}+\lambda ^0,\frac{\lambda ^{{D3a}}}{2}+\lambda ^1,-\frac{\lambda
   ^{{D3a}}}{2},-\frac{\lambda ^{{D3a}}}{2},\frac{\lambda ^{{D3a}}}{2}+\lambda ^{{D3b}}, \right. \nonumber \\
   && \left. \frac{\lambda
   ^{{D3a}}}{2}+\lambda _{{D3b}},\frac{\lambda ^{{D3a}}}{2}+\lambda ^6,\frac{\lambda ^{{D3a}}}{2}+\lambda
   ^7,\frac{\lambda ^{{D3a}}}{2}+\lambda ^8,\frac{\lambda ^{{D3a}}}{2}+\lambda ^9\right\} \;.
\end{eqnarray}
Obvious symmetries of $D1$-$D5$ black hole imply
\begin{equation}
 \lambda'^2 = \lambda'^3 = \lambda'^4 = \lambda'^5 \;,
\end{equation}
and therefore 
\begin{equation}
 \lambda^{D3a} = - \lambda^{D3b} \;. 
\end{equation}
This is the constraint one gets for $D3$-$D3$ solution, which can be verified from the 
explicit solution.
These conclusions can be made just from duality and not using any explicit form of the fields.

\subsubsection*{Relations among $P$s and $T_\phi$}
Like in M-theory case duality method can be used to interpret S and T-duality
relation $\sum_i c_i \lambda^i = 0$ as implying a
relation among the components of the energy-momentum tensor
$T_{A B}$.
Using equation (\ref{strrln}) in equation (\ref{t22bhstring}) one finds
\begin{equation}
 - \frac{p+1}{7-p} = \frac{\frac{1}{8}  (P_R + \sum P_\beta)-P_\perp}{\frac{1}{8}  (P_R + \sum P_\beta)-P_\parallel} \;,
\end{equation}
which on simplification gives 
\begin{equation}\label{STstring}
 P_\parallel+(7-q)P_\perp = P_0 + P_R + m P_a \;.
\end{equation}
Similarly use of equation (\ref{strrln}) in equation (\ref{tphi})
gives 
\begin{equation}\label{STdilaton}
 T_\phi =- \frac{3-p}{2} z \left(P_0+P_R+m P_a)-(8-q)P_\perp\right)
\end{equation}

$\theta^a$ directions are also transverse to brane, so one would expect
$P_\perp = P_a$.
Again it is also natural to take $P_0 = P_\parallel \;$ since $x^0 = t
\;$ is one of the worldvolume coordinates and may naturally be
taken to be on the same footing as the other ones $(x^1, \cdots,
x^p) $. Equation (\ref{STstring}) then implies that $P_\perp = - P_R
\;$. The relation between $P_\parallel $ and $P_R$ is to be
specified by an equation of state.

For now, however, we take $P_0$ and $P_\parallel$ to be
different.  We assume the
equations of state to be of the form $p_{\alpha (I)} = - (1 -
u^I_\alpha) P_{R(I)} $ where $\alpha = (0, i, a)$, $I = 1,
\cdots, N $, and $u^I_\alpha $ are constants. 
Here we give explicitly $u_\alpha$'s for $D1$ and $D5$
black brane solution. 
\begin{eqnarray} 
D1  &:&  u_\alpha =
(\; \; u_0, \; \; u_\parallel,  \; \; u_\perp, \; \; u_\perp, \; \; u_\perp, 
\; \; u_\perp, \; \; u_\perp, \; \; u_\perp, \; \; u_\perp) 
\nonumber \\
D5  &:&  u_\alpha = 
(\; \; u_0, \; \; u_\parallel, \; \; u_\parallel, \; 
\; \; u_\parallel, \; \; \; u_\parallel, \; \; \; u_\parallel,
\; \; u_\perp, \; \; u_\perp, \; \; u_\perp) 
\label{25Wbh} \;.
\end{eqnarray}
Note here that, 
$u_\parallel = u_0 + u_\perp \;$ which follows from equation
(\ref{STstring}).

Using definition of ${\cal G}^{IJ}$, $l^I$ and $\tau$, equation (\ref{genlI}) 
now becomes 
\begin{equation}\label{c3bh}
l^I_{\tau \tau} = - \; \sum_J {\cal G}^{I J} \; e^{l^J} 
+ u_\perp \; \epsilon \; m (m - 1) \; e^{2 (\Lambda - \sigma)}
\; \; .
\end{equation}
Since we know $u_\alpha$'s, 
using equations (\ref{25Wbh}) and (\ref{defnG}), it is now
straightforward to calculate ${\cal G}^{I J} $ for $N$
intersecting branes.  
Similarly using \ref{STdilaton} and equation of motion \ref{tphi}, dilaton 
can be determined. \footnote{Note here that, in black brane case these equations of state are known, 
but in some more general case they may not be known. These duality relations, on the 
other hand, are completely general and may help to get equations of state.}


\section{Applications in time dependent system} \label{sec:TD}
In this paper, we discuss how the duality method works for string theory 
and exotic branes of M-theory in known case. A possible application could be a time 
dependent system. For example, if we consider exotic branes are uniformly distributed
over its common transverse space then we get cosmological solutions similar to \cite{Kalyana Rama:2007ft} -- \cite{Bhowmick:2010dd}. 
We may take 
the line element as
\begin{equation}
 d s^2 = - d t^2 
+ \sum_{i=1}^8 e^{2 \lambda^i(t)} (d x^i)^2 + e^{2 \lambda(t,\xi,\theta)} d \xi^2 
+ e^{2 \sigma(t,\xi,\theta)}  d \theta^2 \; .
\end{equation}
We take both $\xi$ and $\theta$ as compact.
Our $T^A{}_B$s are taken as $T^0{}_0=-\rho$, $T^i{}_i=P_i$, $T^\xi{}_\xi=P_\Xi$
and $T^\theta{}_\theta = P_\Theta$.
The equations of motion look like
\begin{eqnarray}
&\frac{1}{2} (\Lambda_t^2-\sum_i (\lambda_t^i)^2)+\lambda_t \sigma_t +(\lambda_t+\sigma_t)\Lambda_t 
-e^{-2\sigma} (\sigma_\xi^2+\sigma_{\xi\xi}-\lambda_\xi\sigma_\xi)
-e^{-2\lambda} (\lambda_\theta^2+\lambda_{\theta\theta}-\lambda_\theta\sigma_\theta) = \rho, & \\
& \lambda_{tt}^i + \Lambda_t \lambda_t^i = P_i - \frac{1}{9}(\sum P_i +P_\Xi+P_\Theta -\rho), & \\
& \frac{1}{2} (\Lambda_t^2+\sum_i (\lambda_t^i)^2)+\Lambda_t \sigma_t +\sigma_t^2+ \Lambda_{tt}+\sigma_{tt} = -P_\Xi, & \\
& \frac{1}{2} (\Lambda_t^2+\sum_i (\lambda_t^i)^2)+\Lambda_t \lambda_t +\lambda_t^2+ \Lambda_{tt}+\lambda_{tt} = -P_\Theta \;.&
\end{eqnarray}
Here, use of the method discussed in this paper gives a relation similar to 
equation (\ref{zexo}). Known cases of black branes help us to make a guess about equations of state. We may take them as
\begin{eqnarray}
P_\parallel &=& C \rho \\
P_\Xi &=& f_1(\xi,\theta) P_\perp \\
P_\Theta &=& f_2(\xi,\theta) P_\perp \;.
\end{eqnarray}
Here, $C$ is a constant but $f_1(\xi,\theta)$ and $f_2(\xi,\theta)$ introduce non-geometric features. 
With these equations of state, one can solve equations of motion and get a cosmological 
solution \cite{Bhowmick:2015yy}.
Certainly in this universe, expanding spacelike dimensions 
will not be three. We don't know at this 
stage what the physical significance of this non-geometry of the early universe is.

Another possible application is to use similar construction of 
\cite{Bhowmick:2008cq,Bhowmick:2010dd} in 10 dimensional theory. There, a model with 2 sets of $M2$ branes
and 2 sets of $M5$ branes are considered. A similar analysis can be done in 10 dimensional
type II supergravity with a gas of bound state of $F1$ strings, $D2$, $D4$ and $NS5$ branes and bound state of same anti-branes. To do that analysis
equations (\ref{STstring}) and (\ref{STdilaton}) are essential.
This model also give $3+1$ 
dimensional FRW universe with the rest of the dimensions and dilaton stabilized. 
We have better control over string theory than M-theory, so it may be possible to find equations
of state exactly.

\vspace{4mm}
{\bf Acknowledgements:} 
We would like to thank S. Kalyana Rama for his helpful comments on an
earlier version of the draft.

\appendix
\section{Non-BPS Intersection of Branes} \label{nbps}
Here we give an example to show, 
total energy-momentum tensor of non-BPS intersecting branes is not the sum of that of individual branes.
We consider an almost similar configuration of subsection 
{\bf \ref{sec:22'}},
except now our configuration is non-BPS. In this case let us take two sets of 2 
branes along $(x^1,x^2)$ and $(x^2,x^3)$. It is a non-BPS intersection of 
branes. We take $x^1,x^2$ and $x^3$ as compact. Both sets are 
electrically charged. So the 3-form potential, in this case, are
$C_{012} = f_2(r)$ and $C_{023} = f_{2'}(r)$.
So corresponding fields are 
\begin{eqnarray}
 F_{012r} = f_2'(r) = E_2(r), \;\;\;\;\;\; F_{023r} = f_{2'}'(r) =E_{2'}(r) \;. 
\end{eqnarray}
For metric, we may start with an ansatz like (\ref{bh25anz}), 
but it turns out that this ansatz is inconsistent. The reason is explained below. 
Because of the first term in the expression of the energy-momentum tensor 
(equation (\ref{bhTAB})), $T_{13}$ is non-zero, 
\begin{equation}
 T_{13} = \frac{1}{12} g^{00} g^{22} g^{rr} F_{102r} F_{302r} \times 3! \;.
\end{equation}
But since the metric is diagonal, $(R_{13} - \frac{1}{2} g_{13} \, R)$ is zero.
So obviously Einstein equations are not satisfied. Therefore, one has to take a 
different ansatz, the simplest one is almost diagonal metric with only $g_{13}$ non-zero.
That is,
\begin{equation}
\label{nbpsanz}
 ds^2 = -e^{2\lambda^0(r)}\,dt^2+\sum_{i=1}^3 e^{2\lambda^i(r)}\,(dx^i)^2
      +2 \, e^{2\lambda^c} dx^1 dx^3 +
      + e^{2\lambda(r)} \left(dr^2+r^2 d\Omega^2_6 \right) .
\end{equation}
With this ansatz, it turns out non-vanishing components of Einstein tensor
are all diagonal components and $G_{13}$. So now we can equate $G_{MN}$ and
$T_{MN}$. So we will take  the above line element as our ansatz.

We calculate here $T^{M}{}_N$. The non-zero components of
them turn out to be
\begin{eqnarray}
\label{nbpsTab}
 T^0{}_0 &=& \frac{1}{4}\left(g^{00}g^{11}g^{22}g^{rr}F_{012r}F_{012r} 
             + g^{00}g^{22}g^{33}g^{rr}F_{023r}F_{023r}
             + g^{00}g^{13}g^{22}g^{rr}F_{012r}F_{032r} \right) \nonumber \\
 T^1{}_1 &=& \frac{1}{4}\left(g^{00}g^{11}g^{22}g^{rr}F_{012r}F^{012r} 
             - g^{00}g^{22}g^{33}g^{rr}F_{023r}F_{023r}
             + g^{00}g^{13}g^{22}g^{rr}F_{012r}F_{032r} \right) \nonumber \\
 T^2{}_2 &=& \frac{1}{4}\left(g^{00}g^{11}g^{22}g^{rr}F_{012r}F^{012r} 
             + g^{00}g^{22}g^{33}g^{rr}F_{023r}F_{023r}
             + g^{00}g^{13}g^{22}g^{rr}F_{012r}F_{032r} \right) \nonumber \\
 T^3{}_3 &=& \frac{1}{4}\left(-g^{00}g^{11}g^{22}g^{rr}F_{012r}F_{012r} 
             + g^{00}g^{22}g^{33}g^{rr}F_{023r}F_{023r}
             + g^{00}g^{13}g^{22}g^{rr}F_{012r}F_{032r} \right) \nonumber \\
 T^r{}_r &=& \frac{1}{4}\left(g^{00}g^{11}g^{22}g^{rr}F_{012r}F_{012r} 
             + g^{00}g^{22}g^{33}g^{rr}F_{023r}F_{023r}
             + g^{00}g^{13}g^{22}g^{rr}F_{012r}F_{032r} \right) \nonumber \\
 T^a{}_a &=& -\frac{1}{4}\left(g^{00}g^{11}g^{22}g^{rr}F_{012r}F_{012r} 
             + g^{00}g^{22}g^{33}g^{rr}F_{023r}F_{023r}
             + g^{00}g^{13}g^{22}g^{rr}F_{012r}F_{032r} \right) \;, 
\end{eqnarray}
where index $a$ denotes coordinates in $\Omega_6$. There is another component
$T_{13}$. Now because of $T^1{}_3 = g^{11} T_{13} + g^{13} T_{33}$ and 
$T^3{}_1 = g^{33} T_{13} + g^{13} T_{11}$, $T^1{}_3$ and $T^3{}_1$ are not 
symmetric. They turned out to be various combination of $F_{012r}$ and 
$F_{023r}$, and are non-zero.
From above equations, one can see clearly that, total energy-momentum 
tensor is not just the sum of individual energy-momentum tensors
created by each set of branes separately. For example in equations 
(\ref{nbpsTab}) first two terms in each equation give energy-momentum
tensor for individual brane configuration, but the third term is extra.
Besides, $T_{13}$ is a new component, which was not in case of single
$M2$ brane system. So we may conclude that, for non-BPS intersections,
equation (\ref{tabI}) is not satisfied.


\begin{thebibliography}{999}
 \bibitem{Tseytlin:1996bh}
A.~A.~Tseytlin,
Nucl.\ Phys.\  B {\bf 475}, 149 (1996), 
arXiv: hep-th/9604035; 

\bibitem{Tseytlin:1996hi}
A.~A.~Tseytlin,
Nucl.\ Phys.\  B {\bf 487}, 141 (1997), 
arXiv: hep-th/9609212. 

\bibitem{Horowitz:1996ay}
G.~T.~Horowitz, J.~M.~Maldacena and A.~Strominger, 
Phys.\ Lett.\  B {\bf 383}, 151 (1996), 
arXiv: hep-th/9603109. 

\bibitem{Horowitz:1996ac}
G.~T.~Horowitz, D.~A.~Lowe and J.~M.~Maldacena, 
Phys.\ Rev.\ Lett.\  {\bf 77}, 430 (1996), 
arXiv: hep-th/9603195; 

\bibitem{Danielsson:2001xe}
U.~H.~Danielsson, A.~Guijosa and M.~Kruczenski, 
JHEP {\bf 09}, 011 (2001), 
arXiv: hep-th/0106201;


\bibitem{Danielsson:2002qg}
U.~H.~Danielsson, A.~Guijosa and M.~Kruczenski, 
Rev.\ Mex.\ Fis.\  {\bf 49S2}, 61 (2003),
[arXiv:gr-qc/0204010].

\bibitem{Guijosa:2004zn}
A.~Guijosa, H.~H.~Hernandez Hernandez and H.~A.~Morales Tecotl,
JHEP {\bf 03}, 069 (2004), 
[arXiv:hep-th/0402158].

\bibitem{Saremi:2004pi}
O.~Saremi and A.~W.~Peet, 
Phys.\ Rev.\  D {\bf 70}, 026008 (2004),
[arXiv:hep-th/0403170].

\bibitem{Bergman:2004tz}
O.~Bergman and G.~Lifschytz, 
JHEP {\bf 04}, 060 (2004), 
[arXiv:hep-th/0403189].

\bibitem{KalyanaRama:2004fk}
S.~Kalyana Rama, 
Phys.\ Lett.\  B {\bf 593}, 227 (2004), 
[arXiv:hep-th/0404026].

\bibitem{Lifschytz:2004zn}
G.~Lifschytz, 
JHEP {\bf 09}, 009 (2004), 
[arXiv:hep-th/0405042].

\bibitem{Rama:2004iv}
S.~Kalyana Rama and S.~Siwach, 
Phys.\ Lett.\  B {\bf 596}, 221 (2004), 
[arXiv:hep-th/0405084].

\bibitem{Bhowmick:2007tt}
S.~Bhowmick and S.~Kalyana Rama, 
arXiv:0709.3891 [hep-th].

\bibitem{Chowdhury:2006pk}
B.~D.~Chowdhury and S.~D.~Mathur, \\
Class.\ Quant.\ Grav.\  {\bf 24}, 2689 (2007), 
[arXiv:hep-th/0611330]. 

\bibitem{Mathur:2008ez}
S.~D.~Mathur, \\
J.\ Phys.\ Conf.\ Ser.\  {\bf 140}, 012009 (2008), 
[arXiv:0803.3727 [hep-th]]. 

\bibitem{Mathur:2005zp}
For a review of the fuzz ball picture, see 
S.~D.~Mathur, \\
Fortsch.\ Phys.\  {\bf 53}, 793 (2005), 
[arXiv:hep-th/0502050]; 

\bibitem{Mathur:2005ai}
Also see \\
S.~D.~Mathur, \\
Class.\ Quant.\ Grav.\  {\bf 23}, R115 (2006), 
[arXiv:hep-th/0510180]; \\
and also the references therein. 

\bibitem{Kalyana Rama:2007ft} 
  S.~Kalyana Rama,
  Phys.\ Lett.\ B {\bf 656}, 226 (2007)
  [arXiv:0707.1421 [hep-th]].
\bibitem{Bhowmick:2008cq} 
  S.~Bhowmick, S.~Digal and S.~Kalyana Rama,
  Phys.\ Rev.\ D {\bf 79}, 101901 (2009)
  [arXiv:0810.4049 [hep-th]].
  
\bibitem{Bhowmick:2010dd} 
  S.~Bhowmick and S.~Kalyana Rama,
  Phys.\ Rev.\ D {\bf 82}, 083526 (2010)
  [arXiv:1007.0205 [hep-th]].
\bibitem{Bhowmick:2012dc} 
  S.~Bhowmick,
  arXiv:1201.5712 [hep-th].
 
\bibitem{Elitzur:1997zn} 
  S.~Elitzur, A.~Giveon, D.~Kutasov and E.~Rabinovici,
  Nucl.\ Phys.\ B {\bf 509}, 122 (1998)
  [hep-th/9707217].
  
\bibitem{Blau:1997du} 
  M.~Blau and M.~O'Loughlin,
  Nucl.\ Phys.\ B {\bf 525}, 182 (1998)
  [hep-th/9712047].
  
\bibitem{Hull:1997kb} 
  C.~M.~Hull,
  JHEP {\bf 9807}, 018 (1998)
  [hep-th/9712075].
  
\bibitem{Obers:1997kk} 
  N.~A.~Obers, B.~Pioline and E.~Rabinovici,
  Nucl.\ Phys.\ B {\bf 525}, 163 (1998)
  [hep-th/9712084].

\bibitem{Obers:1998fb} 
  N.~A.~Obers and B.~Pioline,
  Phys.\ Rept.\  {\bf 318}, 113 (1999)
  [hep-th/9809039].
 
\bibitem{deBoer:2010ud} 
  J.~de Boer and M.~Shigemori,
  Phys.\ Rev.\ Lett.\  {\bf 104}, 251603 (2010)
  [arXiv:1004.2521 [hep-th]].

\bibitem{deBoer:2012ma} 
  J.~de Boer and M.~Shigemori,
  Phys.\ Rept.\  {\bf 532}, 65 (2013)
  [arXiv:1209.6056 [hep-th]].
  
\bibitem{Hull:2004in} 
  C.~M.~Hull,
  JHEP {\bf 0510}, 065 (2005)
  [hep-th/0406102].
  
\bibitem{Rama:2011xz} 
  S.~K.~Rama,
  arXiv:1111.1897 [hep-th].

\bibitem{Rama:2013jfa} 
  S.~K.~Rama,
  Phys.\ Rev.\ D {\bf 88}, no. 4, 044007 (2013)
  [arXiv:1304.6537 [hep-th]].

\bibitem{Rama:2013vya} 
  S.~K.~Rama,
  Phys.\ Rev.\ D {\bf 89}, 084019 (2014)
  [arXiv:1312.7762 [hep-th]].

  
\bibitem{Aref'eva:1996fw} 
  I.~Y.~Aref'eva and O.~A.~Rychkov,
  Am.\ Math.\ Soc.\ Transl.\  {\bf 201}, 19 (2000)
  [hep-th/9612236].
  
\bibitem{Argurio:1997gt} 
  R.~Argurio, F.~Englert and L.~Houart,
  Phys.\ Lett.\ B {\bf 398}, 61 (1997)
  [hep-th/9701042].
  
\bibitem{Arefeva:1996tw} 
  I.~Y.~Arefeva, K.~S.~Viswanathan, A.~I.~Volovich and I.~V.~Volovich,
  Nucl.\ Phys.\ Proc.\ Suppl.\  {\bf 56B}, 52 (1997)
  [hep-th/9701092].

\bibitem{Aref'eva:1997pf} 
  I.~Y.~Aref'eva, M.~G.~Ivanov and O.~A.~Rychkov,
  In *Kharkov 1997, Supersymmetry and quantum field theory* 25-41
  [hep-th/9702077].

\bibitem{Aref'eva:1997nz} 
  I.~Y.~Aref'eva, M.~G.~Ivanov and I.~V.~Volovich,
  Phys.\ Lett.\ B {\bf 406}, 44 (1997)
  [hep-th/9702079].

\bibitem{Ohta:1997gw} 
  N.~Ohta,
  Phys.\ Lett.\ B {\bf 403}, 218 (1997)
  [hep-th/9702164].

\bibitem{Gauntlett:1997cv} 
  J.~P.~Gauntlett,
  In *Seoul/Sokcho 1997, Dualities in gauge and string theories* 146-193
  [hep-th/9705011].

\bibitem{Clement:2004ii} 
  G.~Clement, D.~Gal'tsov and C.~Leygnac,
  Phys.\ Rev.\ D {\bf 71}, 084014 (2005)
  [hep-th/0412321].

\bibitem{Chen:2005uw} 
  C.~M.~Chen, D.~V.~Gal'tsov and N.~Ohta,
  Phys.\ Rev.\ D {\bf 72}, 044029 (2005)
  [hep-th/0506216].

\bibitem{Gal'tsov:2005vf} 
  D.~Gal'tsov, S.~Klevtsov, D.~Orlov and G.~Clement,
  Int.\ J.\ Mod.\ Phys.\ A {\bf 21}, 3575 (2006)
  [hep-th/0508070].
  
\bibitem{Papadopoulos:1996uq} 
  G.~Papadopoulos and P.~K.~Townsend,
  Phys.\ Lett.\ B {\bf 380}, 273 (1996)
  [hep-th/9603087].
  
\bibitem{Bhowmick:2015yy}
S.~Bhowmick, 
In preparation.  


\end{thebibliography}
\end{document}